\documentclass[10pt]{article}

\usepackage{amsmath,amsthm,amsfonts,mathrsfs,anysize,fancyhdr,epsfig,mdframed, float,graphicx,umoline,dsfont, framed}
\theoremstyle{plain}
\newtheorem{theorem}{Theorem}

\theoremstyle{definition}

\newtheorem{remark}[theorem]{Remark}

\usepackage{chngcntr}
\usepackage{apptools}

\usepackage{amsmath}                                                        
\usepackage{graphicx}                                                       
\usepackage{subfigure}
\usepackage{enumitem}                                                    
\usepackage{hyperref}
\usepackage{amssymb}                                                        
\usepackage[mathscr]{eucal}                                             
\usepackage{chngcntr}
\counterwithin{table}{section}                                                                                   
\counterwithin{figure}{section}                                                                                   
\usepackage{cancel}                                                             
\usepackage[normalem]{ulem}                                                                 
\usepackage{pstricks}
\usepackage{rotating}
\usepackage{lscape}
\usepackage[paperwidth=8.5in,paperheight=11in,top=1.25in, bottom=1.25in, left=1in, right=1in]{geometry}
\usepackage{mathtools}                                                      
\mathtoolsset{showonlyrefs=true}                                    
\usepackage{fixltx2e,amsmath}                                           
\MakeRobust{\eqref}
\def \caratt {{\mathds{1}}}

\usepackage{mathdots}
\usepackage{amsthm}                                                             
\allowdisplaybreaks   
                                                              
\begin{document}
\title{Efficient Computation of Various Valuation Adjustments Under Local L\'evy Models} 


\author{
  Anastasia Borovykh\thanks{Dipartimento di Matematica, Universit\`a di Bologna, Bologna, Italy.
(\textbf{e-mail}:anastasia.borovykh2@unibo.it)}
  \and Andrea Pascucci\thanks{Dipartimento di Matematica, Universit\`a di
Bologna, Bologna, Italy. (\textbf{e-mail}:andrea.pascucci@unibo.it)} \and Cornelis W. Oosterlee\thanks{Centrum Wiskunde \& Informatica, Amsterdam, The Netherlands and Delft University of Technology, Delft, The Netherlands. (\textbf{e-mail}:c.w.oosterlee@cwi.nl)}
}

\maketitle

\begin{abstract}
Various valuation adjustments, or XVAs, can be written in terms of non-linear PIDEs equivalent to FBSDEs. In this paper we develop a Fourier-based method for solving FBSDEs in order to efficiently and accurately price Bermudan derivatives, including options and swaptions,  with XVA under the flexible dynamics of a local L\'evy model: this framework includes a local volatility function and a local jump measure.  Due to the unavailability of the characteristic function for such processes, we use an asymptotic approximation based on the adjoint formulation of the problem.
\end{abstract}


\section{Introduction}\label{sec1}
After the financial crisis in 2007, it was recognized that Counterparty Credit Risk (CCR) poses a
substantial risk for financial institutions. In 2010 in the Basel III framework an additional
capital charge requirement, called Credit Valuation Adjustment (CVA), was introduced to cover the
risk of losses on a counterparty default event for over-the-counter (OTC) uncollateralized
derivatives. The CVA is the expected loss arising from a default by
the counterparty and can be defined as the difference between the risky value and the current
risk-free value of a derivatives contract. CVA is calculated and hedged in the same way as derivatives by many banks,
therefore having efficient ways of calculating the value and the Greeks of these adjustments is
important. 

One common way of pricing CVA is to use the concept of expected exposure, defined as the mean of the exposure distribution at a future date. Calculating these exposures typically involve computationally
time-consuming Monte Carlo procedures, like nested Monte Carlo schemes or the more efficient
least squares Monte Carlo method (LSM)\cite{LongstaffSchwartz}. Recently the Stochastic Grid
Bundling method (SGBM)\cite{shashi2013} was introduced as an improvement of the standard LSM.
This method was extended to pricing CVA for Bermudan options in \cite{feng14}. Another recently
introduced alternative is the so-called finite-differences Monte Carlo method (FDMC) \cite{degraaf}. The FDMC method uses the scenario generation from the Monte Carlo method combined
with finite-difference option valuation. 

Besides CVA, many other valuation adjustments, collectively called XVA, have been
introduced in option pricing in the recent years, causing a change in the way derivatives contracts are priced. For instance, a companies own credit risk is taken into account with a debt value adjustment (DVA). The DVA is the expected gain that will be experienced by the bank in the event that the bank defaults on its portfolio of derivatives with a counterparty. To reduce the credit risk in a derivatives contract, the parties can include a credit support annex
(CSA), requiring one or both of the parties to post collateral. Valuation of derivatives under CSA
was first done in \cite{piterbarg10}. A margin valuation adjustment (MVA) arises when the parties are required to post an initial margin. In this case the cost of posting the initial margin to the counterparty over the length of the contract is known as MVA. Funding value adjustments (FVA) can be interpreted as a
funding cost or benefit associated to the hedge of market risk of an uncollateralized transaction
through a collateralized market. While there is still a debate going on about whether to include
or exclude this adjustment, see \cite{hullwhite2} and \cite{fvadebate} for an
in-depth overview of the arguments, most dealers now seem to indeed take into account the FVA.
The capital value adjustment (KVA) refers to the cost of funding the additional capital that is
required for derivative trades. This capital acts as a buffer against unexpected losses and thus,
as argued in \cite{green}, has to be included in derivative pricing. 

For pricing in the presence of XVA, one needs to redefine the pricing partial differential equation (PDE) by constructing a hedging portfolio with cashflows that are consistent with the additional funding requirements. This has been done for unilateral CCR in \cite{piterbarg10}, bilateral CCR and XVA in \cite{burgard11} and extended to stochastic rates in \cite{lesniewski16}. This results in a non-linear option valuation PDE. 

Non-linear PDEs can be solved by e.g. finite-difference methods or the LSM for solving the corresponding backward stochastic differential equation (BSDE). In \cite{piterbarg15} an efficient forward simulation algorithm that gives the solution of the non-linear PDE as the optimum over solutions of related but linear PDEs is introduced, with the computational cost being of the same order as one forward Monte Carlo simulation. The downside of these numerical methods is the
computational time that is required to reach an accurate solution. An efficient alternative might
be to use Fourier methods for solving the (non-)linear PDE or related BSDE, such as the COS method, as was introduced in \cite{FangO08}, extended to Bermudan options in \cite{FangO09} and to BSDEs in \cite{ruijter15}. In certain cases the efficiency of these methods is further increased due to the ability to additionally use the
fast Fourier transform (FFT).

In this paper we consider an exponential L\'evy-type model with a state-dependent jump
measure and propose an efficient Fourier-based method to solve for Bermudan derivatives, including options and swaptions, with
XVA. We derive, in the presence of state-dependent jumps, a non-linear partial integro-differential equation (PIDE)
and its corresponding BSDE for an OTC derivative between a bank $B$
and its counterparty $C$ in the presence of CCR, bilateral collateralization, MVA, FVA and KVA, by setting up a hedging portfolio in which we focus on hedging the default risks and take into account the different rates associated with different types of lending. We
extend the Fourier-based method known as the BCOS method, developed in \cite{ruijter15}, to solve
the BSDE under L\'evy models with non-constant coefficients. As this method requires the
knowledge of the characteristic function of the forward process, which, in the case of the L\'evy
process with variable coefficients, is not known, we will use an approximation of the
characteristic function obtained by the adjoint expansion method developed in \cite{pascucci-riga}, \cite{LorigPP2015} and extended to the defaultable L\'evy process with a state-dependent jump measure in \cite{borovykh}. Compared to other state-of-the-art methods for calculating XVAs, like
Monte Carlo methods and PDE solvers, our method is more efficient and/or flexible. The efficiency is both due to the availability of the characteristic function in closed form through the adjoint expansion method and the fast convergence of the COS method.
Furthermore we propose an alternative Fourier-based method for explicitly pricing the
CVA term in case of unilateral CCR for Bermudan derivatives under the local L\'evy model. The advantage of this method is that
is allows us to use the FFT, resulting in a fast and efficient calculation. The Greeks, used for
hedging CVA, can be computed at almost no additional cost.

The rest of the paper is structured as follows. In Section \ref{sec2} we introduce the L\'evy models with non-constant coefficients. In Section \ref{sec3} we derive the non-linear PIDE and corresponding BSDE for pricing contracts under XVA. In Section \ref{sec4} we propose the Fourier-based method for solving this BSDE and in Section \ref{sec51} this method is extended to pricing Bermudan contracts. In Section \ref{sec52} an alternative FFT-based method for pricing and hedging the CVA term is proposed and Section \ref{sec6} presents numerical examples validating the accuracy and efficiency of the proposed methods. 

\section{The model}\label{sec2}
We consider a defaultable asset $S_t$ whose risk-neutral dynamics are given by 
 \begin{align}
 S_t &= \caratt_{\{t<\zeta\}}e^{X_t},\\
 dX_t &= \mu (t,X_t)dt+\sigma (t,X_t)dW_t+\int_\mathbb{R}q d\tilde N_t(t,X_{t-},dq),\\\label{eq:hetmodel}
d\tilde N_t(t,X_{t-},dq) &= dN_t(t,X_{t-},dq)-a(t, X_{t-})\nu (dq)dt,\\
 \zeta &= \inf\{t\geq 0 :\int_0^t\gamma(s,X_{s})ds\geq \varepsilon\},
\end{align}
where $d\tilde N_t(t,X_{t-},dq)$ is a compensated random measure with state-dependent L\'evy
measure $$\nu(t,X_{t-},dq) = a(t, X_{t-})\nu(dq).$$ The
default time $\zeta$ of $S_t$ is defined in a canonical way as the first arrival time of a doubly
stochastic Poisson process with local intensity function $\gamma(t,x)\geq 0$, and $\varepsilon
\sim \mathrm{Exp}(1)$ and is independent of $X_t$. This way of modeling default is also considered
in a diffusive setting in \cite{JDCEV} and for exponential L\'evy models in \cite{capponi}. Thus,
our model includes a local volatility function, a local jump measure, and a default probability
which is dependent on the underlying. We define the filtration at time $t$ of the market observer to be
$\mathcal{G}_t=\mathcal{F}^X_t\vee \mathcal{F}^D_t$, where $\mathcal{F}^X_t$ is the filtration generated
by $X$ upto time $t$ and $\mathcal{F}_t^D:=\sigma(\{\zeta\leq u\},u\leq t)$, for $t\ge0$, is the filtration of
the default. Using this definition of default, the probability of default is
 \begin{align}\label{eq:probdef}
 \text{PD}(t) := \mathbb{P}(\zeta\leq t)= 1-\mathbb{E}\left[e^{-\int_0^t\gamma(s,X_s)ds}\right].
 \end{align}
We assume furthermore
\begin{equation}\label{nusomm}
  \int_\mathbb{R}e^{|q|}a(t,x)\nu(dq)<\infty.
\end{equation}
Imposing that the discounted asset price $\tilde S_t := e^{-rt}S_{t}$ is a
$\mathcal{G}$-martingale under the risk-neutral measure, we get the following restriction on the drift coefficient:
\begin{align}\label{eq:martdrift}
\mu(t,x) = \gamma(t,x)+r-\frac{\sigma^2(t,x)}{2}-a(t,x)\int_\mathbb{R}\nu(dq)(e^q-1-q),
\end{align}
with $r$ being the risk-free (collateralized) rate. In the whole of the paper we assume
deterministic, constant interest rates, while the derivations can easily be extended to time-dependent rates. The integro-differential operator of the
process is given by (see e.g. \cite{Pascucci2011})
 \begin{align}
 L u(t,x) =&\partial_t u(t,x)+\mu(t,x)\partial_xu(t,x) -\gamma(t,x)u(t,x)+\frac{\sigma^2(t,x)}{2}\partial_{xx}u(t,x)\nonumber\\
 &+a(t,x)\int_\mathbb{R}\nu(dq)(u(t,x+q)-u(t,x)-q\partial_x u(t,x)).\label{eq:opL}
 \end{align}

\section{XVA computation}\label{sec3}
Consider a bank $B$ and its counterparty $C$, both of them might default. Assume they enter into a contract paying $\Phi(S_t)$ at maturity. Let $\phi(x) = \Phi(e^x)$, and assume the risk-neutral dynamics of
the underlying as in \eqref{eq:hetmodel} with the drift given by \eqref{eq:martdrift}. 
Define $\hat u(t,x)$ to be the
value to the bank of the (default risky) portfolio with valuation adjustments referred to as XVA and $u(t,x)$ to be the risk-free
value.  Note that the difference between these two values is called the \emph{total valuation adjustment} and in our setting this consists of 
\begin{align}\label{eq:xva}
\textnormal{TVA}:=\hat u(t,x)-u(t,x)=\textnormal{CVA}+\textnormal{DVA}+\textnormal{KVA}+\textnormal{MVA}+\textnormal{FVA}.
\end{align}
The risk-free value $u(t,x)$ solves a linear
PIDE:
\begin{align}\label{eq:linearpdenoxva}
Lu(t,x) &= ru(t,x),\\
u(T,x) &= \phi(x),
\end{align}
where $L$ is given in \eqref{eq:opL}. Assuming the dynamics in
\eqref{eq:hetmodel}, this linear PIDE can be solved with the methods presented in \cite{borovykh}.

\subsection{Derivative pricing under CCR and bilateral CSA agreements}\label{sec31}
In \cite{burgard11}, the authors derive an extension to the Black-Scholes PDE in the presence of a
bilateral counterparty risk in a jump-to-default model with the underlying being a diffusion,
using replication arguments that include the funding costs. In \cite{lesniewski16} this derivation is
extended to a multivariate diffusion setting with stochastic rates in the presence of CCR,
assuming that both parties $B$ and $C$ are subject to default. To mitigate the CCR, both parties
exchange collateral consisting of the initial margin and the variation margin. The parties are obliged to hold regulatory capital, the cost of which is the KVA and
face the costs of funding uncollateralized positions through collateralized markets, known as FVA. Both \cite{burgard11} and \cite{lesniewski16}
extend the approach of \cite{piterbarg10}, in which unilateral collateralization was considered. We extend their approach to
derive the value of $\hat u(t,x)$ when the underlying follows the jump-diffusion defined in
\eqref{eq:hetmodel}. We assume a one-dimensional underlying diffusion and consider all rates to be
deterministic and, for ease of notation, constant. We specify different rates, defined in Table \ref{tab000}, for different types of lending.

\begin{table}[h!]
\label{tab000}
\begin{center}
\begin{tabular}{ l|l|l|l}
Rate & Definition& Rate & Definition\\ \hline\hline
$r$ & \parbox[t]{6cm}{the risk-free rate}&$r_R$& \parbox[t]{6.5cm}{the rate received on funding secured by the underlying asset}\\
$r_D$& \parbox[t]{6cm}{the dividend rate in case the stock pays dividends}&$r_F$& \parbox[t]{6.5cm}{the rate
received on unsecured funding}\\
$r_B$ & \parbox[t]{6cm}{the yield on a bond of the bank $B$ }&$r_C$ & \parbox[t]{6cm}{the yield on the bond of the counterparty $C$}\\
$\lambda_B$&$\lambda_B := r_B-r$ & $\lambda_C$ & $\lambda_C := r_C-r$ \\
$\lambda_F$&$\lambda_F := r_F-r$&$R_B$ & \parbox[t]{6cm}{ the recovery rate of the bank}\\
$R_C$ &  \parbox[t]{6.5cm}{the recovery rate of the counterparty}&&\\
\hline\hline
\end{tabular}
\end{center}
   \caption{Definitions of the rates used throughout the paper.}
\end{table}

Assume that the parties $B$ and $C$ enter into a derivative contract on the spot
asset that pays the bank $B$ the amount $\phi(X_T)$ at maturity $T$. The value of this derivative
to the bank at time t is denoted by $\hat u(t,x,\mathcal{J}^B,\mathcal{J}^C)$ and depends on the value of the underlying
$X$ and the default states $\mathcal{J}^B$ and $\mathcal{J}^C$ of the bank $B$ and counterparty $C$, respectively. Define $I^{TC}$ to be the initial margin posted by the bank to the counterparty, $I^{FC}$ the initial margin posted by the counterparty to the bank and $I^V(t)$ to be the variation margin on which a rate $r_I$ is paid or received. The initial margin is constant throughout the duration of the contract. Let $K(t)$ be the regulatory capital on which a rate of $r_K$ is paid/received.

The cashflows are viewed from the perspective of the bank $B$. At the default time of either the counterparty or the bank, the value of the derivative to the bank $\hat u(t,x)$ is determined with a mark-to-market rule $M$, which may be equal to either the derivative value $\hat u(t,x,0,0)$ prior to default or the risk-free derivative value $u(t,x)$, depending on the specifications in the ISDA master agreement. Denote by $\tau^B$ and $\tau^C$ the random default times of the bank and the counterparty respectively. We will use the notation $x^+=\max(x,0)$ and $x^-=\min(x,0)$. In a situation in which the counterparty defaults, the bank is already in the possession of $I^V+I^{FC}$. If the outstanding value $M-(I^V+I^{FC})$ is negative, the bank has to pay the full amount $(M-I^V-I^{FC})^-$, while if the contract has a positive value to the bank, it will recover only $R_C(M-I^V-I^{FC})^+$. Using a similar argument in case the bank defaults, we find the following boundary conditions:
\begin{align}
\theta^B_t &:=\hat u(t,x,1,0)= I^V(t)-I^{TC}+(M-I^V(t)+I^{TC})^++R^B(M-I^V(t)+I^{TC})^-,\\
\theta^C_t &:=\hat u(t,x,0,1)= I^V(t)+I^{FC}+R^C(M-I^V(t)-I^{FC})^++(M-I^V(t)-I^{FC})^-,
\end{align}
so that the portfolio value at default is given by
$$\theta_\tau = 1_{\tau^C<\tau^B}\theta_\tau^C+1_{\tau^B<\tau^C}\theta_\tau^B,$$
with $\tau = \min(\tau^B,\tau^C)$.
Further we introduce the default risky, zero-recovery, zero-coupon bonds (ZCBs) $P^B$ and $P^C$ with respective maturities $T^B$ and $T^C$ with face value one if the issuer has not defaulted, and zero otherwise. Assume the dynamics for $P^B_t$ and $P_t^C$ to be given by $P_t^B=\caratt_{\{\tau^B>t\}}e^{r_Bt}$ and $P_t^C=\caratt_{\{\tau^C>t\}}e^{r_Ct}$, so that
\begin{align}
dP_t^B &= r_BP^B_tdt-P_{t-}^Bd\mathcal{J}_t^B,\\
dP_t^C &= r_CP^C_tdt-P_{t-}^Cd\mathcal{J}_t^C,
\end{align}
with $\mathcal{J}_t^B=\caratt_{\tau^B\leq t}$ and $\mathcal{J}_t^C=\caratt_{\tau^C\leq t}$, where the default times $\tau^B$ and $\tau^C$ are defined in a canonical way as the first arrival time of a doubly stochastic Poisson process with intensity functions $\gamma^B$ and $\gamma^C$, respectively (see also the definition of the defaultable asset in \eqref{eq:hetmodel}). 
We define the market interest rates for $B$ and $C$ to be $r_B=r+\gamma^B$ and $r_C=r+\gamma^C$, so that by the usual arguments (see, for instance, \cite[Section 2.2]{linetsky2006bankruptcy}) the discounted bonds $e^{-rt}P_t^B$ and $e^{-rt}P_t^C$ are martingales under the risk-neutral measure. 

We construct a hedging portfolio consisting of the shorted derivative, $\alpha_C$ units of $P^C$, $\alpha_B$ units of $P^B$ and $g$ units of cash:
$$\Pi(t) = -\hat u(t,x) + \alpha_B(t)P^B_t+\alpha_C(t)P^C_t+g(t).$$
In other words, since we assume both the underlying asset process and the tradeable bonds $P_B$ and $P_C$ to be risk-neutral, we focus on hedging the risk arising from the defaults of both $B$ and $C$ by means of the default-risky bonds. 

If the value of the derivative is positive to $B$, it will incur a cost at the counterparties' default. To hedge this, $B$ shorts $P^C$, i.e. $\alpha_C\leq 0$. If we assume $B$ can borrow the bond close to the risk-free rate $r$ (i.e. no haircut) through a repurchase agreement, it will incur financing costs of $r\alpha_C(t)P_t^Cdt$. 
The cashflows from the collateralization follow from the rate $r_{TC}$ received and $r_{FC}$ paid on the initial margin and the rate $r_I$ paid or received on the collateral, depending on whether $I^V>0$, and the bank receives collateral, or $I^V<0$, and the bank pays collateral respectively. From holding the regulatory capital we incur a cost of $r_KK(t)$. Finally, the rates $r$ and $r_F$ are respectively received or paid on the surplus cash in the account. This cash consists of the gap between the shorted derivative value and the collateral and the cost of buying $\alpha_B$ bonds $P^B$ in order for $B$ to hedge its own default, i.e. $-\hat u(t,x)-I^V(t)+I^{TC}-\alpha_B(t)P_t^B$. Thus, the total change in the cash account is given by 
\begin{align}
dg(t) =& [-r\alpha_C(t)P^C_t+r_{TC}I_{TC}-r_{FC}I_{FC}-r_II^V(t)-r_KK(t)\\
&+r(-\hat u(t,x)-I^V(t)+I_{TC}-\alpha_B(t)P^B_t)+\lambda_F(-\hat u(t,x)-I^V(t)+I_{TC}-\alpha_B(t)P^B_t)^-]dt.
\end{align}
Note that this is in contrast with the change in cash in a portfolio without the XVA arising from the different types of funding, i.e. where we assume the cash in the portfolio simply earns the risk-free rate
\begin{align}
dg(t) = -r \hat u(t,x)dt.
\end{align}
Assuming the portfolio is self-financing we have 
\begin{align}
d\Pi(t)=& -d\hat u(t,x)+\alpha_B(t)dP^B_t+\alpha_C(t)dP^C_t+dg(t).
\end{align}
Applying It\^o's Lemma to $\hat u(t,x)$ gives us:
\begin{align}
d\hat u(t,x) =& L\hat u(t,x)dt +\sigma(t,x)\partial_x\hat u(t,x)dW_t+\int_\mathbb{R}(\hat u(t,x+q)-\hat u(t,x))d\tilde N(t,x,dq)\\
&-(\theta^B-\hat u(t,x))d\mathcal{J}^B_t-(\theta^C-\hat u(t,x))d\mathcal{J}^C_t,
\end{align}
with the operator $L$ as in \eqref{eq:opL}.
Thus, we find,
\begin{align}
d\Pi =& -L\hat u(t,x)dt -\sigma(t,x)\partial_x\hat u(t,x)dW_t-\int_\mathbb{R}(\hat u(t,x+q)-\hat u(t,x))d\tilde N(t,x,dq)\\
&+(\theta^B-\hat u(t,x))d\mathcal{J}^B_t+(\theta^C-\hat u(t,x))d\mathcal{J}^C_t-\alpha^B(t)P_{t-}^Bd\mathcal{J}_t^B-\alpha^C(t)P_{t-}^Cd\mathcal{J}_t^C\\
&+[\alpha^B(t)\lambda_BP_t^B+\alpha^C(t)\lambda_CP^C_t+(r_{TC}+r)I^{TC}-r_{FC}I^{FC}-(r_I+r)I^V(t)\\
&-r_KK(t)+r\hat u(t,x)+\lambda_F(-\hat u(t,x)-I^V(t)+I^{TC}-\alpha^B(t)P^B_t)^-]dt.
\end{align}
By choosing
\begin{align}
\alpha_B = -\frac{\theta^B - \hat u(t,x)}{P_B}, \;\;\; \alpha_C = -\frac{\theta^C - \hat u(t,x)}{P_C},
\end{align}
we hedge the jump-to-default risk in the hedging portfolio, i.e.,
\begin{align}
d\Pi =& -L\hat u(t,x)dt +\sigma(t,x)\partial_x\hat u(t,x)dW_t-\int_\mathbb{R}(\hat u(t,x+q)-\hat u(t,x))d\tilde N(t,X_{t-},dq)\\
&+[-(\theta^B-\hat u(t,x))\lambda_B-(\theta^C-\hat u(t,x))\lambda_C+(r_{TC}+r)I^{TC}-r_{FC}I^{FC}-(r_I+r)I^V(t)\\
&-r_KK(t)+r\hat u(t,x)+\lambda_F(\theta^B-I^V(t)+I^{TC})^-]dt.
\end{align}
Then, using the fact that the portfolio has to satisfy the martingale condition in the risk-neutral world, i.e. $\mathbb{E}[d\Pi] = 0$, we find the non-linear pricing PIDE to be 
\begin{align}\label{eq:thepde}
L\hat u(t,x) =& f(t,x,\hat u(t,x)),
\end{align}
where we have defined
\begin{align}
f(t,x,\hat u(t,x))=&-(\theta^B(t)-\hat u(t,x))\lambda_B-(\theta^C(t)-\hat u(t,x))\lambda_C+(r_{TC}+r)I^{TC}-r_{FC}I^{FC}\\
&-(r_I+r)I^V(t)-r_KK(t)+r\hat u(t,x)+\lambda_F(\theta^B-I^V(t)+I^{TC})^-.
\end{align}

\subsection{BSDE representation}\label{sec32}
In this section we will cast the PIDE in \eqref{eq:thepde} in the form of a Backward Stochastic Differential Equation. In the methods where we make use of BSDEs we assume $\gamma(t,x)=0$. We begin by recalling the non-linear Feynman-Kac theorem in the presence of jumps, see Theorem 4.2.1 in \cite{delong13}.

\begin{theorem}[Non-linear Feynman-Kac Theorem]\label{theorem1}
Consider $X_t$ as in \eqref{eq:hetmodel}. We assume $\mu$, $\sigma$ and $a$ to be Lipschitz continuous in $x$ and additionally $|a(t,x)|\leq K$. 
Consider the BSDE
\begin{align}
Y_t =\ & \phi(X_T) + \int_t^T
f\left(s,X_s,Y_s,Z_s,a(s,X_{s-})\int_\mathbb{R}V_s(q)\delta(q)\nu(dq)\right)ds-\int_t^TZ_sdW_s\\\label{eq:setfbsdes1}
&-\int_t^T\int_\mathbb{R}V_s (q)d\tilde   N_s(s,X_s,q),
\end{align}
where the generator $f$ is continuous and satisfies the Lipschitz condition in the space variables, $\delta$ is a measurable, bounded function and the terminal condition $\phi(x)$ is measurable and Lipschitz continuous. 
Consider the non-linear PIDE
\begin{align}\label{eq:bsdepde}
\begin{cases}
 Lu(t,x) = f(t,x,u(t,x),\partial_xu(t,x)\sigma(t,x),a(t,x)\int_\mathbb{R}(u(t,x+q)-u(t,x))\delta(q)\nu(dq)),\\
 u(T,x) = \psi(x).
\end{cases}
\end{align}
If the PIDE in \eqref{eq:bsdepde} has a solution $u(t,x)\in C^{1,2}$, the FBSDE in \eqref{eq:setfbsdes1} has a unique solution $(Y_t,Z_t,V_t(q))$ that can be represented as
\begin{align}
&Y_s^{t,x} = u(s,X_s^{t,x}),\\ 
&Z_s^{t,x} = \partial_xu(s,X_s^{t,x})\sigma(s,X_s^{t,x}),\\
&V_s^{t,x}(q) = u(s,X_s^{t,x}+q)-u(s,X_s^{t,x}),\qquad q\in\mathbb{R},
\end{align}
for all $s\in[t,T]$, where $Y$ is a continuous, real-valued and adapted process and where the control processes $Z$ and $V$ are continuous, real-valued and predictable.
\end{theorem}
In our case, the BSDE corresponding to the PIDE in \eqref{eq:thepde} reads
\begin{align}\label{eq:fundbsde}
Y_t = \phi(X_T) + \int_t^T f(s,X_s,Y_s)ds-\int_t^TZ_sdW_s-\int_t^T\int_\mathbb{R}V_s(q)d\tilde N(s,X_s,dq),
\end{align}
where we have defined the driver function to be
\begin{align}
f(t,x,y) =& -\lambda_B(\theta^B-y)-\lambda_C(\theta^C-y)+(r_{TC}+r)I^{TC}-r_{FC}I^{FC}-(r_I+r)I^V(t)\\
&-r_KK(t)+ry+\lambda_F(\theta^B-I^V(t)+I^{TC})^-.
\end{align}

\subsection{A simplified driver function}\label{sec323}
Following \cite{green}, one can derive that the KVA is a function of trade properties (i.e. maturity, strike) and/or the exposure at default, which in turn is a function of the portfolio value, so that the cost of holding the capital can be rewritten as $r_KK(t)=r_Kc_1\hat u(t,x),$ with $c_1$ being a function of the trade properties. The collateral is paid when the portfolio has a negative value, and received when the collateral has a positive value. Assuming the collateral is a multiple of the portfolio value we have $I^V(t)= c_2\hat u(t,x)$, where $c_2$ is some constant. Then, the driver function is simply a function of the portfolio value.

\begin{remark}
Note that in the case of `no collateralization' or `perfect collateralization', the driver function reduces to $f(t,\hat u(t,x)) = r_u(t)\max(\hat u(t,x),0)$, for a function $r_u$ here left unspecified. In this case the BSDE is similar to the one considered in \cite{piterbarg15}.
\end{remark}

\section{Solving FBSDEs}\label{sec4}
In this section we extend the BCOS method from \cite{ruijter15} to solving FBSDEs under local L\'evy
models with variable coefficients and jumps (without default, i.e. $\gamma(t,x)=0$). The conditional expectations resulting from the
discretization of the FBSDE are approximated using the COS method. This requires the
characteristic function, which we approximate using the {Adjoint Expansion Method} of
\cite{pascucci-riga} and \cite{borovykh}.

\subsection{Discretization of the BSDE}\label{sec41}
Consider the forward process $X_t$ as in \eqref{eq:hetmodel} and the BSDE $Y_t$ as in \eqref{eq:fundbsde} with a more general driver function $f(t,x,y,z)$. Define a partition $0=t_0<t_1<...<t_N=T$ of $[0,T]$ with a fixed time step $\Delta t = t_{n+1}-t_n$, for $n=N-1,...0$. Rewriting the set of FBSDEs we find,
\begin{align}
X_{n+1}=&X_n+\int_{t_n}^{t_{n+1}}\mu(s,X_s)ds+\int_{t_n}^{t_{n+1}}\sigma(s,X_s)dW_s+\int_{t_n}^{t_{n+1}}\int_\mathbb{R}qd\tilde N_s(s,X_{s-},dq),\\ \label{eq:discry}
Y_{n}=&Y_{n+1}+\int_{t_n}^{t_{n+1}}f\left(s,X_s,Y_s,Z_s\right)ds-\int_{t_n}^{t_{n+1}}Z_sdW_s-\int_{t_n}^{t_{n+1}}\int_\mathbb{R}V_s(q)d\tilde N_s(s,X_{s-},dq).
\end{align}
One can obtain an approximation of the process $Y_t$ by taking conditional expectations with respect to the underlying filtration $\mathcal{G}_n$, using the independence of $W_t$ and $\tilde N_t(t,X_{t-},dq)$ and by approximating the integrals that appear with a theta-method, as first done in \cite{zhao12} and extended to BSDEs with jumps in \cite{ruijter15}:
\begin{align}
Y_n &\approx  \mathbb{E}_n[Y_{n+1}]+\Delta t \theta_1f\left(t_n,X_n,Y_n,Z_n\right)+\Delta t (1-\theta_1)\mathbb{E}_n\left[f\left(t_{n+1},X_{n+1},Y_{n+1},Z_{n+1}\right)\right].
\end{align}
Let $\Delta W_s:=W_s-W_n$ for $t_n\leq s\leq t_{n+1}$. Multiplying both sides of equation \eqref{eq:discry} by $\Delta W_{n+1}$, taking conditional expectations and applying the theta-method gives
\begin{align}
Z_n &\approx -\theta_2^{-1}(1-\theta_2)\mathbb{E}_n[Z_{n+1}]+\frac{1}{\Delta t}\theta_2^{-1}\mathbb{E}_n[Y_{n+1}\Delta W_{n+1}]\\
&+\theta_2^{-1}(1-\theta_2)\mathbb{E}_n\left[f\left(t_{n+1},X_{n+1},Y_{n+1},Z_{n+1}\right)\Delta W_{n+1}\right].
\end{align}
Since in our scheme the terminal values are functions of time $t$ and the Markov process $X$, it is easily seen that there exist deterministic functions $y(t_n,x)$ and $z(t_n,x)$ so that
\begin{align}
Y_n=y(t_n,X_n), \;\;\; Z_n=z(t_n,X_n).
\end{align}
The functions $y(t_n,x)$ and $z(t_n,x)$ are obtained in a backward manner using the following scheme
\begin{align}\label{eq:scheme1}
y(t_N,x)=&\phi(x), \;\;\; z(t_N,x) = \partial_x\phi(x)\sigma(t_N,x),\\
&\textnormal{for $n=N-1,...,0$:}\\\label{eq:scheme2}
y(t_n,x) =&  \mathbb{E}_n[y(t_{n+1},X_{n+1})]+\Delta t \theta_1f\left(t_n,x\right)+\Delta t (1-\theta_1)\mathbb{E}_n\left[f(t_{n+1},X_{n+1})\right],\\\label{eq:scheme3}
z(t_n,x) =& -\frac{1-\theta_2}{\theta_2}\mathbb{E}_n[z(t_{n+1},X_{n+1})]+\frac{1}{\Delta t}\theta_2^{-1}\mathbb{E}_n[y(t_{n+1},X_{n+1})\Delta W_{n+1}]\\
&+\frac{1-\theta_2}{\theta_2}\mathbb{E}_n\left[f(t_{n+1},X_{n+1})\Delta W_{n+1}\right],
\end{align}
where we have simplified notations with
\begin{align}
f(t,X_t) := f\left(t,X_t,y(t,X_t),z(t,X_t)\right).
\end{align}
In the case $\theta_1>0$ we obtain an implicit dependence on $y(t_n,x)$ in \eqref{eq:scheme2} and we use $P$ Picard iterations starting with initial guess $\mathbb{E}_n[y(t_{n+1},X_{n+1})]$ to determine $y(t_n,x)$. 

\subsection{The characteristic function}\label{sec42}
Is it well-known (see, for instance, \cite[Section 2.2]{linetsky2006bankruptcy}) that the risk-free pre-default price
$u(t,x)$ of a European option on the defaultable asset $S_t$ with maturity $T$ and payoff $\phi(X_{T})$  is given by
\begin{align}\label{e1}
u(t,x) = \caratt_{\{\zeta>t\}} e^{-r(T-t)}\mathbb{E} \left[e^{-\int_t^T \gamma(s,X_s) ds}\phi(X_T)  | X_t \right],\;\;\; t\leq T,
\end{align}
in the measure corresponding to the dynamics in \eqref{eq:hetmodel}. 
Thus, in order to compute the price of an option, we must evaluate functions of the form
\begin{align}\label{expectation}
 v(t,x):= \mathbb{E} \left[e^{-\int_t^T \gamma(s,X_s) ds}\phi(X_T)| X_t = x \right] .
\end{align}
Under standard assumptions, by the Feynman-Kac theorem, $v$ can be expressed
as the classical solution of the following Cauchy problem
\begin{align}\label{eq:v.pide}
&\begin{cases}
 L v(t,x)=0,\qquad & t\in[0,T[,\ x\in\mathbb{R}, \\
 v(T,x) =  \phi(x),& x \in\mathbb{R},
\end{cases}
\end{align}
with $L$ as in \eqref{eq:opL}.

The function $v$ in \eqref{expectation} can be represented as an integral with respect to the
transition distribution of the defaultable log-price process $\log S_t$:
\begin{align}\label{eq:v.def1}
 v(t,x) = \int_\mathbb{R} \phi(y)\Gamma(t,x;T,dy),
\end{align}
where $\Gamma(t,x;T,dy)$ is the Green's function of the PIDE in \eqref{eq:v.pide} and we say that its Fourier transform
 $$\hat\Gamma(t,x;T,\xi):=\mathcal{F}(\Gamma(t,x;T,\cdot))(\xi):= \int_\mathbb{R}e^{i\xi y}\Gamma(t,x;T,dy),\qquad \xi\in\mathbb{R},$$
is the characteristic function of $\log S$. Following \cite{pascucci-riga} and \cite{borovykh} we
expand the state-dependent coefficients
 $$s(t,x):=\frac{\sigma^2(t,x)}{2},\qquad \mu(t,x), \qquad \gamma(t,x),\qquad a(t,x),$$
around some point $\bar{x}$. The coefficients $s(t,x)$, $\gamma(t,x)$ and $a(t,x)$ are
assumed to be continuously differentiable with respect to $x$ up to order $n\in\mathbb{N}$. 

Introduce the $n$th-order approximation of $L$ in
\eqref{eq:opL}:
\begin{align}
 L_n =&\ L_0 + \sum_{k=1}^n\Big( (x-\bar x)^k\mu_k(t)+(x-\bar x)^k s_k(t) \partial_{xx}-(x-\bar x)^k\gamma_k(t)\\
 &+\int_\mathbb{R}(x-\bar x)^ka_k(t)\nu(dq)(e^{q\partial_x}-1-q\partial_x)\Big),
 \end{align}
where
  \begin{align}
 L_0 &=  \partial_t +\mu_0(t)\partial_x+ s_0(t) \partial_{xx}-\gamma_0(t)
  +\int_\mathbb{R}a_0(t)\nu(dq)(e^{q\partial_x}-1-q\partial_x),
 \end{align}
and
\begin{align}
  s_k= \frac{\partial_x^k s(\cdot,\bar x)}{k!},\qquad
  \gamma_k = \frac{\partial_x^k \gamma(\cdot,\bar x)}{k!},\qquad
  \mu_k(dq) = \frac{\partial_x^k \mu (\cdot,\bar x)}{k!},\qquad a_k= \frac{\partial_x^k a(\cdot,\bar x)}{k!}\qquad\ k\ge 0.
 \end{align}
The basepoint $\bar x$ is a constant parameter which can be chosen freely. In general the simplest
choice is $\bar x = x$ (the value of the underlying at initial time $t$).

Assume for a moment that $L_{0}$ has a fundamental solution $G^{0}(t,x;T,y)$ that
is defined as the solution of the Cauchy problem
 $$\begin{cases}
 L_0 G^{0}(t,x;T,y) =0\qquad & t\in[0,T[,\ x\in\mathbb{R}, \\
 G^{0}(T,\cdot;T,y) =\delta_{y}.
 \end{cases}$$
In this case we define the $n$th-order approximation of $\Gamma$ as
 $$\Gamma^{(n)}(t,x;T,y) = \sum_{k=0}^n G^{k}(t,x;T,y),$$
where, for any $k\ge 1$ and $(T,y)$, $G^{k}(\cdot,\cdot;T,y)$ is defined recursively through the
following Cauchy problem
 $$
  \begin{cases}
 L_0 G^{k}(t,x;T,y) = -\sum\limits_{h=1}^k(L_h-L_{h-1})G^{k-h}(t,x;T,y)\qquad & t\in[0,T[,\ x\in\mathbb{R}, \\
 G^{k}(T,x;T,y) =0,& x \in\mathbb{R}.
 \end{cases}
 $$
Notice that
 \begin{align}
 L_k-L_{k-1}=& (x-\bar x)^k \mu_h(t) \partial_{x} +(x-\bar x)^k s_k(t) \partial_{xx}-(x-\bar x)^k\gamma_k(t)\\
 &+\int_\mathbb{R}(x-\bar x)^ka_k(t)\nu(dq)(e^{q\partial_x}-1-q\partial_x).
 \end{align}
Correspondingly, the $n$th-order approximation of the characteristic function $\hat \Gamma$ is defined to
be
\begin{equation}\label{adapprox}
 \hat \Gamma^{(n)}(t,x;T,\xi)=\sum_{k=0}^n \mathcal{F}\left(G^{k}(t,x;T,\cdot)\right)(\xi):=\sum_{k=0}^n \hat G^{k}(t,x;T,\xi),\qquad \xi\in\mathbb{R}.
\end{equation}
Now, by transforming the simplified Cauchy problems into adjoint problems and solving these in the Fourier space we find
\begin{align}\label{eq:Ghat}
 \hat G^{0}(t,x;T,\xi) &= e^{i\xi x}e^{\int_t^T\psi(s,\xi)ds},\\
 \hat G^{k}(t,x;T,\xi) &= -\int_t^Te^{\int_s^T\psi(\tau,\xi)d\tau}\mathcal{F}\left(\sum_{h=1}^k\left(\tilde
  L_h^{(s,\cdot)}(s)-\tilde L_{h-1}^{(s,\cdot)}(s)\right)G^{k-h}(t,x;s,\cdot)\right)(\xi)ds,
\end{align}
with
\begin{align}
 \psi(t,\xi) = i\xi\mu_0(t)
 +s_0(t)\xi^2+\int_\mathbb{R}a_0\nu(t,dq)(e^{iz\xi}-1-iz\xi),
 \end{align}
 \begin{align}
 \tilde L_h^{(t,y)}(t)-\tilde L_{h-1}^{(t,y)}(t) &= \mu_h(t)h(y-\bar x)^{h-1}+\mu_h(t)(y-\bar x)^h\partial_y-\gamma_h(t)(y-\bar x)^h\\
 & +s_h(t) h(h-1)(y-\bar x)^{h-2}+s_h(t) (y-\bar x)^{h-1} \left(2h\partial_y+(y-\bar x)\partial_{yy}\right)\\
 &+\int_\mathbb{R}a_h(t)\bar\nu(dq)\left((y+q-\bar x)^he^{q\partial_y}-(y-\bar x)^h-q\left(h(y-\bar x)^{h-1}-(y-\bar
 x)^h\partial_y\right)\right),
\end{align}
where $\bar \nu(dq) = \nu(-dq)$.
\begin{remark}\label{r3}
After some algebraic manipulations it can be shown, see \cite{borovykh}, that the characteristic function approximation of order $n$ is a function
of the form
\begin{equation}\label{eq:struc3}
  \hat\Gamma^{(n)}(t,x;T,\xi):= e^{i\xi x} \sum_{k=0}^n (x-\bar x)^k g_{n,k}(t,T,\xi),
\end{equation}
where the coefficients $g_{n,k}$, with $0\leq k\leq n$, depend only on $t,T$ and $\xi$, but not on
$x$. The approximation formula can thus always
be split into a sum of products of functions depending only on $\xi$ and functions that are linear combinations of $(x-\bar x)^m e^{i\xi x}$, $m\in\mathbb{N}_{0}$.
\end{remark}

\begin{remark}[Error estimates for the approximated characteristic function]
Similar to the derivation in \cite{borovykh}, one can derive the error bounds for the characteristic function approximation. Let $n=0,1$ and assume the coefficients $s(t,x)$, $\gamma(t,x)$ and $a(t,x)$ are continuously differentiable with bounded derivatives up to order $n$. For the $n$th-order approximation $\Gamma^{(n)}(t,x;T,\xi)$, for any $\bar x\in \mathbb{R}$,
\begin{align}
\left|\Gamma(t,x;T,\xi)-\Gamma^{(n)}(t,x;T,\xi)\right|\leq C(T,\xi)((T-t)^2+(T-t)(x-\bar x))^{\frac{n+1}{2}}.
\end{align}
Note that if $\bar x = x$, the bound reduces to $C(T,\xi)(T-t)^{n+1}$. 
\end{remark}

\subsection{The COS formulae}\label{sec43}
The conditional expectations are approximated using the COS
method, which was developed in \cite{FangO09} and applied to FBSDEs with jumps in \cite{ruijter15}.
The conditional expectations arising in the equations \eqref{eq:scheme2}-\eqref{eq:scheme3} are all of the form $\mathbb{E}_n[h(t_{n+1},X_{n+1})]$ or $\mathbb{E}_n[h(t_{n+1},X_{n+1})\Delta W_{n+1}]$.  The COS formula for the first type of conditional expectation reads
\begin{align}
&\mathbb{E}_n^x[h(t_{n+1},X_{n+1})]\approx \sideset{}{'}\sum_{j=0}^{J-1}H_j(t_{n+1})\textnormal{Re}\left(\hat\Gamma\left(t_n,x;t_{n+1},\frac{j\pi}{b-a}\right)\exp\left(ij\pi\frac{-a}{b-a}\right)\right),
\end{align}
where $\sideset{}{'}\sum$ denotes an ordinary summation with the first term weighted by one-half, $J>0$ is the number of Fourier-cosine coefficients we use, $H_j(t_{n+1})$ denotes the $j$th Fourier-cosine coefficients of the function $h(t_{n+1},x)$ and $\hat\Gamma\left(t_n,x;t_{n+1},\xi \right)$ is the conditional characteristic function of the process $X_{n+1}$ given $X_n=x$.
For the second type of conditional expectation, using integration by parts, we obtain
\begin{align}
\mathbb{E}_n^x&[h(t_{n+1},X_{n+1})\Delta W_n]\\
&\approx \Delta t\sigma(t_n,x)\sideset{}{'}\sum_{j=0}^{J-1}H_j(t_{n+1})\textnormal{Re}\left(i\frac{j\pi}{b-a}\hat\Gamma\left(t_n,x;t_{n+1},\frac{j\pi}{b-a}\right)\exp\left(ij\pi\frac{-a}{b-a}\right)\right).
\end{align}
See \cite{ruijter15} for the full derivations.
\begin{remark}
Note that these formulas are obtained by using an Euler approximation of the forward process and using the 2nd-order approximation of the characteristic function of the actual process. We have found this to be more exact than using the characteristic function of the Euler process, which is equivalent to using just the 0th-order approximation of the characteristic function.
\end{remark}

Finally we need to approximate the Fourier-cosine coefficients $H_j(t_{n+1})$ of $h(t_{n+1},x)$ at time points $t_n$, where $n=0,...,N$.
The Fourier-cosine coefficient of $h$ at time $t_{n+1}$ is defined by
\begin{align}\label{eq:fouriercos}
&H_j(t_{n+1})=\frac{2}{b-a}\int_a^bh(t_{n+1},x)\cos\left(j\pi\frac{x-a}{b-a}\right)dx.
\end{align}
Due to the structure of the approximated characteristic function of the local L\'evy process, see
\eqref{eq:struc3}, the coefficients of the functions $z(t_{n+1},x)$ and the
explicit part of $y(t_{n+1},x)$ can be computed using the FFT algorithm, as we do in Appendix
\ref{app1}, because of the matrix in \eqref{eq:integraal} being of a certain form with constant diagonals. In order to
determine $F_j(t_{n+1})$, the Fourier-Cosine coefficient of the function
$$f\left(t_{n+1},x,y(t_{n+1},x),z(t_{n+1},x)\right),$$
due to the intricate dependence on the functions $z$ and $y$ we choose to approximate the integral in $F_j$ by a discrete Fourier-Cosine transform (DCT). For
the DCT we compute the integrand, and thus the functions $z(t_{n+1},x)$ and
$y(t_{n+1},x)$, on an equidistant $x$-grid. Note that in this case we can easily
approximate \emph{all} Fourier-Cosine coefficients with a DCT (instead of the FFT). If we take $J$ grid points defined by 
$x_i:=a+(i+\frac{1}{2})\frac{b-a}{J}$ and $\Delta x = \frac{b-a}{J}$ we find, using the mid-point
integration rule, the approximation
\begin{align}
H_j(t_{n+1})\approx \frac{2}{J}\sideset{}{'}\sum_{i=0}^{J-1}h(t_{n+1},x_i)\cos\left(j\pi \frac{2i+1}{2J}\right),
\end{align}
which can be calculated using the DCT algorithm, with a computational complexity of $O(J\log J)$.
\begin{remark}
We define the truncation range $[a,b]$ as follows:
\begin{align}\label{eq:truncrange}
[a,b]:=\left[c_1-L\sqrt{c_2+\sqrt{c_4}},c_1+L\sqrt{c_2+\sqrt{c_4}}\right],
\end{align} where $c_n$ is the
$n$th cumulant of log-price process $\log S$, as proposed in \cite{FangO08}. The cumulants are
calculated using the 0th-order approximation of the characteristic function.
\end{remark}

\section{XVA computation for Bermudan derivatives}\label{sec5}
The method in Section \ref{sec4} allows us to compute the XVA as in \eqref{eq:xva}, consisting of CVA, DVA, MVA, KVA and FVA. In this section, we apply this method to computing Bermudan derivative values with XVA. The resulting method -- the solution of the non-linear XVA PDE through a BSDE-type method -- is an efficient alternative to finite-difference methods as well as to the Monte-Carlo based method developed in \cite{piterbarg15}. The efficiency is both due to the availability of the characteristic function in closed form through the adjoint expansion method and the fast convergence of the COS method. Furthermore, in finite difference methods complications may arise in the implementation of the scheme for jump diffusions. Since our proposed method works in the Fourier space, the jump component is easily handled by means of an additional term in the characteristic function and does not cause any further difficulties. 

For the CVA component in the XVA we develop an alternative method, which due to the ability of the FFT, results in a particularly efficient computation.

\subsection{XVA computation}\label{sec51}
Consider an OTC derivative contract between the bank $B$ and the counterparty $C$ on the underlying asset $S_t$ given by \eqref{eq:hetmodel} with $\gamma(t,x)=0$ with a Bermudan-type exercise possibility: there is a finite set of so-called exercise moments $\{t_1,...,t_{M}\}$ prior to the maturity,
with $0\le t_{1}<t_2<\cdots<t_{M}=T$. The payoff from the point-of-view of bank $B$ is given by $\phi(t_m, X_{t_m})$.
Denote $\hat u(t,x)$ to be the risky Bermudan option value and $c(t,x)$ the continuation value. By the dynamic programming approach, the value for a Bermudan derivative with XVA and $M$ exercise dates $t_1,...,t_M$ can be expressed by a backward recursion as
\begin{align}\hat u(t_{M},x)=\phi(t_M,x),\end{align}
and the continuation value solves the non-linear PIDE defined in \eqref{eq:thepde}
\begin{align}\label{eq:bermud2}
 \begin{cases}
\begin{cases}
Lc(t,x) =f(t,x,c(t,x)),\qquad \;\; t\in[t_{m-1},t_{m}[\\
c(t_m,x) = \hat u(t_m,x)
\end{cases}\\
 \hat u(t_{m-1},x)=\max\{\Phi(t_{m-1},x),c(t_{m-1},x)\}, \;\;m\in\{2,\dots,M\}.
\end{cases}
\end{align}
The derivative value is set to be $\hat u(t,x)=c(t,x)$ for $t\in]t_{m-1},t_m[$, and, if $t_1>0$, also for $t\in[0,t_1[$.
The payoff function might take on various forms:
\begin{enumerate}
\item (Portfolio) Following \cite{piterbarg15}, we can consider $X_t$ to be the process of a portfolio which can take on both positive and negative values. Then, when exercised at time $t_m$, bank $B$ receives the portfolio so that $\phi(t_m, x) =e^x$.
\item (Bermudan option) In case the Bermudan contract is an option, the option value to the bank can not have a negative value for the bank. At the same time, in case of default of the bank itself, the counterparty loses nothing. In this case the framework simplifies to one with unilateral collateralization and default risk and the payoff at time $t_m$, if exercised, is given by $\phi(t_m,x)=(K-e^x)^+$ for a put and $\phi(t_m,x)=(e^x-K)^+$ for a call with $K$ being the strike price.
\item (Bermudan swaptions) A Bermudan swaption is an option in which the holder, bank $B$, has the right to exercise and enter into an underlying swap with fixed end date $t_{M+1}$. If the swaption is exercised at time $t_m$ the underlying swap starts with payment dates $\mathcal{T}_m=\{t_{m+1},...,t_{M+1}\}$. Working under the forward measure corresponding to the last reset date $t_M$, the payoff function is given by
\begin{align}
\phi(t_m,x) = N^S\left(\sum_{k=m}^M\frac{P(t_m,t_{k+1},x)}{P(t_m,t_M)}\Delta t\right)\max(c_p(S(t_m,\mathcal{T}_m,x)-K),0),
\end{align}
where $N^S$ is the notional, $c_p=1$ for a payer swaption and $c_p=-1$ for a receiver swaption, $P(t_m,t_k,x)$ is the price of a ZCB conditional on $X_{t_m}=x$ and $S(t_m,\mathcal{T}_m,x)$ is the forward swap rate given by
\begin{align}
S(t_m,\mathcal{T}_m,x)=\left(1-\frac{P(t_m,t_{m+1},x)}{P(t_m,t_M,x)}\right)\big /\left(\sum_{k=m}^M\frac{P(t_m,t_{k+1},x)}{P(t_m,t_M,x)}\Delta t\right ).
\end{align}
\end{enumerate}
To solve for the continuation value we define a partition with $N$ steps $t_{m-1}=t_{0,m}<t_{1,m}<t_{2,m}<...<t_{n,m}<...<t_{N,m}=t_m$ between two exercise dates $t_{m-1}$ and $t_m$, with fixed time step $\Delta t_n :=t_{n+1,m}-t_{n,m}$. Applying the method developed in Section \ref{sec4}, we find the following time iteration for the continuation value:
\begin{align}\label{eq:formulaym}
\textnormal{At time $t_{N,m}$ set:}&\\
c(t_{N,m},x)&=\hat u(t_m,x)\\
\textnormal{for }n=N-1,...,&0 \textnormal{ compute:}\\
c(t_{n,m},x)&\approx \Delta t_{n} \theta_1f(t_{n,m},x,c(t_{n,m},x))+\sideset{}{'}\sum_{j=0}^{J-1}\Psi_j(x)(C_j(t_{n+1,m})+\Delta t_n(1-\theta_1)F_j(t_{n+1,m})),\label{eq:formulay}
\end{align}
where we have defined
\begin{align}
&\Psi_j(x) =\textnormal{Re}\left(\hat\Gamma\left(t_{n,m},x;t_{n+1,m},\frac{j\pi}{b-a}\right)\exp\left(ij\pi\frac{-a}{b-a}\right)\right),
\end{align}
and the Fourier-cosine coefficients are given by
\begin{align}
&C_j(t_{n+1,m})=\frac{2}{b-a}\int_a^bc(t_{n+1,m},x)\cos\left(j\pi\frac{x-a}{b-a}\right)dx,\\
&F_j(t_{n+1,m})=\frac{2}{b-a}\int_a^bf(t_{n+1,m},x,c(t_{n+1,m},x))\cos\left(j\pi\frac{x-a}{b-a}\right)dx.
\end{align}
In order to determine the function $c(t_n,x)$, we will perform $P$ Picard iterations. To evaluate
the coefficients with a DCT we need to compute the integrands $c(t_{n+1,m},x)$ and $f(t_{n+1,m},x,c(t_{n+1,m},x))$ on the equidistant $x$-grid with $x_i$, for $i=0,...,J-1$. In order to
compute this at each time step $t_{n,m}$ we thus need to evaluate $c(t_{n,m},x)$ on the $x$-grid with $J$ equidistant points using formula \eqref{eq:formulay}.
The matrix-vector product in the formula results in a computational time of order $O(J^2)$.  

\begin{remark}[Convergence of the Picard iterations]
A Picard iteration is used to find the fixed-point $c$ of $c = \Delta t\theta_1f(t_{n,m},x,c)+h(t_{n,m},x),$
where $f(t,x,c)$ and $h(t,x)$ are respectively the implicit and explicit parts of the equation. Due to the computational domain of $c(t,x)$ being bounded by $[a,b]$, we can thus say that $f(t,x,c(t,x))$ is also bounded. If the driver function $f(t,x,c)$ is Lipschitz continuous in $c$, i.e. $\exists$ $L^{Lipz}$ such that $|f(t,x,c_1)-f(t,x,c_2)|\leq L^{Lipz}|c_1-c_2|$, and $\Delta t_n$ is small enough such that $\Delta t \theta_1 L^{Lipz}<1$, a unique fixed-point exists and the Picard iterations converge towards that point for any initial guess. In particular, for the XVA case the non-linearity is of the form $f(t,x,c) = -r\max(c,0)$, and this is Lipschitz continuous with $L^{Lipz}=1$. Thus for $\Delta t$ sufficiently small, the Picard iteration converges to a unique fixed-point.
\end{remark}

The total algorithm for computing the value of a Bermudan contract with XVA can be summarised as in Algorithm 1 in Figure \ref{fig0}. 
The total computational time for the algorithm is of order
\begin{align}\label{eq:complex1}
O(M\cdot N(J + J^2+PJ+J\log_2 J)),
\end{align} 
consisting of the computation for $M\cdot N$ times the computation of the characteristic function on the $x$-grid (due to the availability of the analytical approximation) of $O(J)$, computation of the matrix-vector multiplications in the formulas for $c(t_{n,m},x)$ and $z(t_{n,m},x)$ of $O(J^2)$, initialization of the Picard method with $\mathbb{E}_n[c(t_{n+1},X_{n+1}]$ in $O(J^2)$ operations, 
computation of the $P$ Picard approximations for $c(t_{n,m},x)$ in $O(PJ)$ and computing the Fourier coefficients $F_j(t_n)$ and $C_j(t_n)$ with the DCT in $O(J\log_2 J)$ operations. 

\begin{figure}[h]\label{fig0}
\begin{framed}
\begin{enumerate}
\item Define the $x$-grid with $J$ grid points given by $x_i=a+(i+\frac{1}{2})\frac{b-a}{J}$ for $i=0,...,J-1$.
\item Calculate the final exercise date values $c(t_{N,M},x)=\hat u(t_M,x)$ on the $x$-grid and compute the terminal coefficients $C_j(t_M)$ and $F_j(t_M)$ using the DCT.
\item Recursively for the exercise dates $m = M-1,...,0$ do:
\begin{enumerate}
\item For time steps $n=N-1,...,0$ do:
\begin{enumerate}
\item Compute $c(t_{n,m},x)$ using formula \eqref{eq:formulay} and use this to determine $f(t_{n,m},x,c(t_{n,m},x))$ on the $x$-grid.
\item Subsequently, use these to determine $F_j(t_{n,m})$ and $C_j(t_{n,m})$ using the DCT.
\end{enumerate}
\item Compute the new terminal condition $c(t_{N,m-1},x)=\max\{\phi(t_{0,m},x),c(t_{0,m},x)\}$ (either analytically or numerically) and the corresponding Fourier-cosine coefficient.
\end{enumerate}
\item Finally $\hat u(t_0,x_0) = c(t_{0,0},x_0)$.
\end{enumerate}
\end{framed}
\caption{Algorithm 1: Bermudan derivative valuation with XVA}
\end{figure}

\subsection{An alternative for CVA computation}\label{sec52}
In this section we present an efficient alternative way of calculating the CVA term in \eqref{eq:xva} in the case of unilateral CCR using a
Fourier-based method. Due to the ability of using the FFT this method is considerably faster for computing the CVA than the method presented in Section \ref{sec51}. We use the definition of CVA at time $t$ given by $$\textnormal{CVA}(t) =
\hat u(t,X_t)- u(t,X_t),$$ where $u(t,X_t)$ is as usual the default-free value of the Bermudan option ($\gamma(t,x)=0$),
while $\hat u(t,X_t)$ is the value including default ($\gamma(t,x)\neq 0$). We consider the model as defined in
\eqref{eq:hetmodel}. We will compute $u(t,X_t)$ and $\hat u(t,X_t)$ using the COS method and the
approximation of the characteristic function (as derived in Section \ref{sec43}), without default
and with default, respectively. In case of a default the payoff becomes zero. Note that the risky option value $\hat u(t,x)$ computed with the characteristic function for a defaultable underlying corresponds exactly to the option value in which the counterparty might default, with the probablity of default, $PD(t)$, defined as in \eqref{eq:probdef}. Thus, in this case we have unilateral CCR and $\zeta = \tau_C$, the default time of the counterparty. 

Using the definition of the defaultable $S_t$, it is well-known (see, for instance, \cite[Section 2.2]{linetsky2006bankruptcy}) that the risky no-arbitrage value of the Bermudan option on the defaultable asset $S_t$ at time $t$ is given by
\begin{align}\hat u\left(t,X_{t}\right)=\caratt_{\{\zeta>t\}}\sup_{\tau \in \{t_1,...,t_M\}}\mathbb{E}\left[e^{-\int_{t}^{\tau}
\left(r+\gamma(s,X_s)\right) ds}\phi(\tau,X_{\tau})|X_{t}\right].
\end{align}
\begin{remark}[Wrong-way risk]
By allowing the dependence of the default intensity on the underlying, a simplified form of wrong-way risk is already incorporated into the CVA valuation.
\end{remark}
For a Bermudan put option with strike price $K$, we simply have
$\phi(t,x)=\left(K-x\right)^{+}$. 
By the dynamic programming approach, the option value
can be expressed by a backward recursion as
 \begin{align}
 \hat u(t_{M},x)=\caratt_{\{\zeta>t_{M}\}}\max(\phi(t_{M},x),0),
 \end{align}
and
\begin{align}
 c(t,x)=
 \mathbb{E}\left[
 e^{\int_{t}^{t_m}\left(r+\gamma(s,X_s)\right)ds}\hat u(t_{m},X_{t_{m}})|X_{t}=x\right],\qquad &t\in[t_{m-1},t_{m}[\\
 \hat u(t_{m-1},x)=\caratt_{\{\zeta>t_{m-1}\}}\max\{\phi(t_{m-1},x),c(t_{m-1},x)\},\qquad
 &m\in\{2,\dots,M\}.\label{eq:bermud}
\end{align}
Thus to find the risky option price $\hat u(t,X_t)$ one uses the defaultable asset with $\gamma(t,x)$ representing the default intensity of the counterparty and in order to get the default-free value $u(t,X_t)$ one uses the default-free asset by setting $\gamma(t,x)=0$. The CVA adjustment is calculated as the difference between the two. Both $\hat u(t,x)$ and $u(t,x)$ are calculated using the approximated characteristic function and the COS method applied to the continuation value \cite{borovykh}. Due to the characteristic function being of the form \eqref{eq:struc3}, we are able to use the FFT in the matrix-vector multiplication when computing the continuation values of the Bermudan option with and without default, reducing this operation from $O(J^2)$ to $O(J\log_2 J)$. For more details, we refer to Appendix \ref{app1}. The total complexity of the calculation of the CVA value for a Bermudan option with $M$ exercise dates is then $O(M J \log_2 J)$. Comparing this to \eqref{eq:complex1}, in which the most time-consuming operations were indeed the matrix-vector products of order $O(J^2)$ that resulted from the computation of the functions on the $x$-grid of size $J$, we conclude that the method for CVA computation is indeed significantly faster due to the ability of using the FFT.

\subsubsection{Hedging CVA}\label{sec521}
In practice CVA is hedged and thus practitioners require efficient ways to compute the sensitivity of the CVA with respect to the underlying.  The widely used bump- and revalue- method, while resulting in precise calculations, might be slow to compute. Using the Fourier-based approach we find explicit formulas allowing for an easy computation of the first- and second-order derivatives of the CVA with respect to the underlying. For the first-order and second-order Greeks we have
\begin{align}
\Delta  &=\ e^{-r(t_{1}-t_0)}\sideset{}{'}\sum_{j=0}^{J-1}\textnormal{Re}\left(e^{ij\pi\frac{x-a}{b-a}}\left(\frac{ij\pi}{b-a}g_{n,0}^d
 \left(t_0,t_{1},\frac{j\pi}{b-a}\right)+g_{n,1}^d\left(t_0,t_{1},\frac{j\pi}{b-a}\right)\right)\right) V_j^d(t_1)\\
 &-\ e^{-r(t_{1}-t_0)}\sideset{}{'}\sum_{j=0}^{J-1}\textnormal{Re}\left(e^{ij\pi\frac{x-a}{b-a}}\left(\frac{ij\pi}{b-a}g_{n,0}^r
 \left(t_0,t_{1},\frac{j\pi}{b-a}\right)+g_{n,1}^r\left(t_0,t_{1},\frac{j\pi}{b-a}\right)\right)\right) V_j^r(t_1),
 \end{align}
 \begin{align}
\frac{\partial \Delta}{\partial X}  &=\ e^{-r(t_{1}-t_0)}\sideset{}{'}\sum_{j=0}^{J-1}\textnormal{Re}\bigg(e^{ij\pi\frac{x-a}{b-a}}\bigg(-\frac{ij\pi}{b-a}
 g_{n,0}^d\left(t_0,t_{1},\frac{j\pi}{b-a}\right)-g_{n,1}^d\left(t_0,t_{1},\frac{j\pi}{b-a}\right)\\
 &+\ 2\frac{ij\pi}{b-a}g_{n,1}^d\left(t_0,t_{1},\frac{j\pi}{b-a}\right)
 +\left(\frac{ij\pi}{b-a}\right)^2g_{n,0}^d\left(t_0,t_{1},\frac{j\pi}{b-a}\right)+2g_{n,2}^d\left(t_0,t_{1},\frac{j\pi}{b-a}\right)\bigg)\bigg) V_j^d(t_1)\\
 &-\ e^{-r(t_{1}-t_0)}\sideset{}{'}\sum_{j=0}^{J-1}\textnormal{Re}\bigg(e^{ij\pi\frac{x-a}{b-a}}\bigg(-\frac{ij\pi}{b-a}
 g_{n,0}^r\left(t_0,t_{1},\frac{j\pi}{b-a}\right)-g_{n,1}^r\left(t_0,t_{1},\frac{j\pi}{b-a}\right)\\
&-\ 2\frac{ij\pi}{b-a}g_{n,1}^r\left(t_0,t_{1},\frac{j\pi}{b-a}\right)
 +\left(\frac{ij\pi}{b-a}\right)^2g_{n,0}^r\left(t_0,t_{1},\frac{j\pi}{b-a}\right)+2g_{n,2}^r\left(t_0,t_{1},\frac{j\pi}{b-a}\right)\bigg)\bigg) V_j(t_1)^r,
\end{align}
where $V_k^d$ and $V_k^r$ are the Fourier-cosine coefficients with the defaultable and default-free characteristic function terms, $g_{n,h}^d$ and $g_{n,h}^r$, respectively.

\section{Numerical experiments}\label{sec6}
In this section we present numerical examples to justify the accuracy of the methods in practice. We compute the XVA with the method presented in Section \ref{sec51} and the CVA in the case of unilateral CCR with the method from  Section \ref{sec52}, which we show is more efficient for cases in which one only needs to compute the CVA. We compare the results of solving the BSDE with the COS method and the adjoint expansion of the characteristic function to the values obtained by using a least-squares Monte-Carlo method for computing the conditional expected values in the BSDE as done in e.g. \cite{bender12}. 

The computer used in the experiments has an Intel Core i7 CPU with a 2.2 GHz processor. We use the
second-order approximation of the characteristic function. We have found this to be sufficiently
accurate by numerical experiments and theoretical error estimates. The formulas for the
second-order approximation are simple, making the methods easy to implement.

\subsection{A numerical example for XVA}\label{sec61}
Here, we check the accuracy of the method from Section \ref{sec51}. We will compute the Bermudan option value with XVA using a simplified driver function given by $f(t,\hat u(t,x)) = -r\max(\hat u(t,x),0)$. Our method is easily extendible to the driver function in Section \ref{sec32}. Consider $X_t$ to be a portfolio process and the payoff, if exercised at time $t_m$, to be given by $\Phi(t_m,x) =
x$. In this case the value we can receive at every exercise date is the value of the
portfolio. Consider the model in Section \ref{sec2} without default, with a local jump measure and a local
volatility function with CEV-like dynamics and Gaussian jumps defined by
\begin{align}\label{eq:parameters1}
&\sigma(x) = be^{\beta x},\\ \label{eq:parameters2}
& \nu(x,dq) = \lambda e^{\beta x}
\frac{1}{\sqrt{2\pi\delta^2}}\exp\left(\frac{-(q-m)^2}{2\delta^2}\right)dq.
\end{align}
We assume the following parameters in equations \eqref{eq:parameters1}-\eqref{eq:parameters2}, unless otherwise mentioned: $b=0.15$, $\beta = -2$, $\lambda = 0.2$, $\delta = 0.2$, $m=-0.2$, $r=0.1$, $K=1$ and $X_0=0$ (so that $S_0 = 1$). In the LSM the number of time steps is taken to be 100 and we simulate $10^5$ paths. In the COS method we take $J=256$, $\theta_1=0.5$ and $N=10$, $M=10$, making the total number of time steps $N\cdot M = 100$.  The truncation range is determined as in \eqref{eq:truncrange} with $L=10$. Due to the state-dependent coefficients in the underlying dynamics in \eqref{eq:parameters1}-\eqref{eq:parameters2} we use the approximated characteristic function as derived in Section \ref{sec42} with the second-order approximation, i.e. $\hat \Gamma^{(2)}(t,x;T,\xi)$ and take $\bar x = x$, where $x=\{x_i\}_{i=0}^{J-1}$. Note that we thus compute the values, including those of the characteristic function, on the complete $x$-grid. In the final iteration when computing $\hat u(t_0,X_0)$ we use $\bar x = X_0$. 

\begin{figure}[H]
\begin{center}
\includegraphics[scale = 0.3]{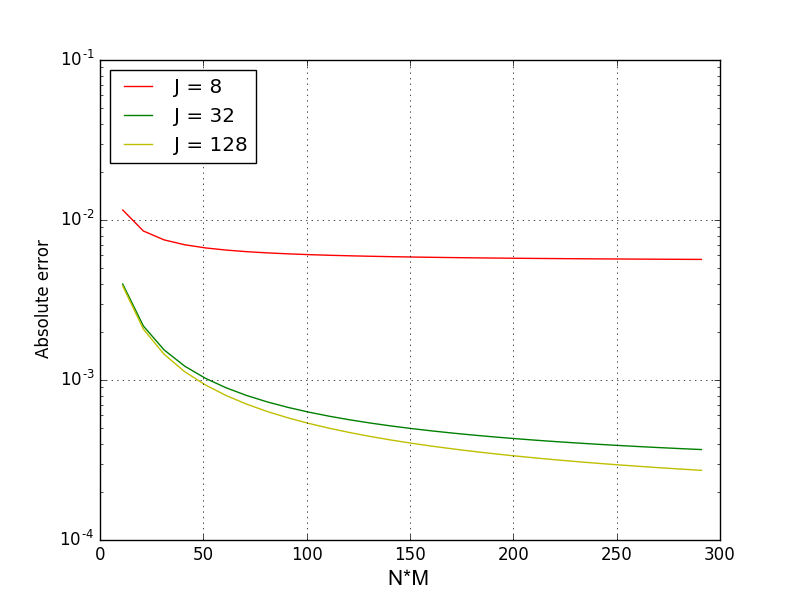}
\includegraphics[scale = 0.3]{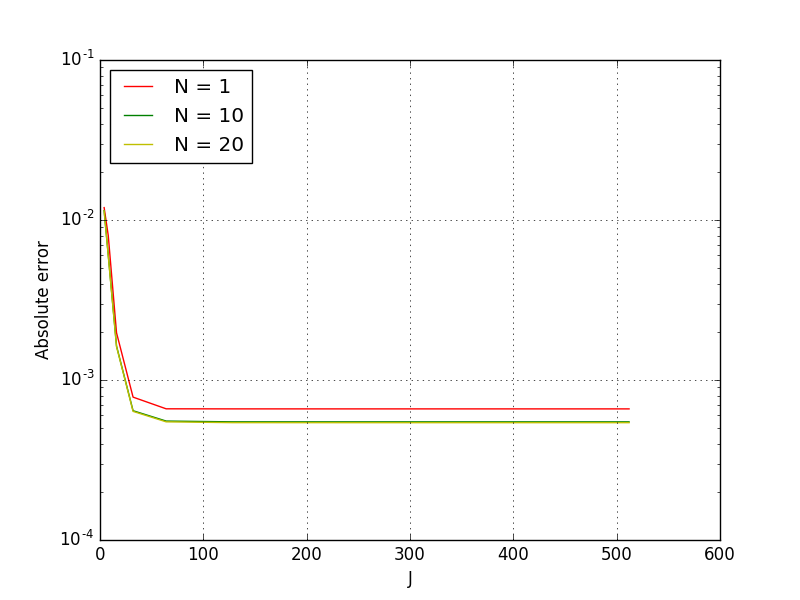}
\end{center}
   \caption{Convergence of the upper bound of the 95\% confidence interval of the absolute error in the COS approximation $\hat u(0,X_0)$ with $S_0 = 0.4$ compared to a LSM for varying parameters $J$ and $N$.}
\label{figconv}
\end{figure}

In Table \ref{taberr} we analyse the error in the approximation of $\hat u(t_0,X_0)$ with $S_0=0.4$ for different values of the discretization parameter $N$ and the number of grid points (and Fourier-cosine coefficients) $J$. We compare the approximated COS value to the 95\% confidence interval obtained by a LSM. Accurate results are quickly obtained for small values of both $J$ and $N$. In Figure \ref{figconv} we plot the upper bound of the 95\% confidence interval of the absolute error in the approximation for varying $J$ and $N$. We observe approximately a linear convergence and note that the error stops decreasing at some point for increasing values of $J$ and $N$. This can be due to the error being dominated by the approximated characteristic function. In particular we observe that $J=32$ and $N=10$ seem to be sufficient parameters to achieve a satisfactory accuracy in the approximation.

The results for $\hat u(t_0,X_0)$ of the COS approximation method compared to a 95\% confidence interval of the value obtained through a LSM are presented in Table \ref{tab1}. These results show
that our method is able to solve non-linear PIDEs accurately. The CPU time of the approximating method depends on the
number of time steps $M\cdot N$ and is approximately $5\cdot (N\cdot M)$ ms.

\begin{table}[H]
\begin{center}
\begin{tabular}{ c||c|c|c|c}
&$N=1$&$N=10$&$N=20$&$N=30$\\\hline\hline
$J=8$&6.4E-03$-$6.9E-03&4.3E-03$-$4.8E-03&4.9E-03$-$5.3E-03&5.3E-03$-$5.8E-03\\
$J=16$&2.3E-03$-$2.7E-03&8.8E-04$-$1.3E-03&6.2E-04$-$1.1E-03&5.4E-04$-$9.2E-04\\
$J=32$&1.7E-03$-$2.0E-03&4.2E-04$-$8.3E-04&2.4E-04$-$6.3E-04&1.6E04$-$5.8E-04\\
$J=64$&1.4E-03$-$1.9E-03&2.2E-04$-$6.5E-04&1.6E-04$-$2.3E-04&1.2E-04$-$2.9E-04\\
$J=128$&1.7E-04$-$6.0E-04&2.1E-04$-$6.6E-04&2.3E-04$-$6.5E-04&1.9E-04$-$6.1E-04\\
$J=256$&2.1E-04$-$6.6E-04&3.7E-04$-$7.7E-04&1.5E-04$-$5.7E-04&1.2E-04$-$3.1E-04\\\hline\hline
\end{tabular}
\label{taberr}
\end{center}
   \caption{The 95\% confidence interval of the absolute error in the COS approximation of $\hat u(0,X_0)$ with $S_0 = 0.4$ compared to a LSM for varying parameters $J$ and $N$.}
\end{table}

\begin{table}[h]
\begin{center}
\begin{tabular}{ l|l|l|l }
maturity $T$&$S_0$&MC value with XVA&COS value with XVA\\\hline \hline
0.5&0&0.03770$-$0.03838&0.03809\\
&0.2&0.2326$-$0.2330&0.2320\\
&0.4&0.4251$-$0.4254&0.4243\\
&0.6&0.6169$-$0.6171&0.6158\\
&0.8&0.8077$-$0.8079&0.8069\\
&1&1.000$-$1.000&1.0000\\ \hline \hline
1&0&0.07374$-$0.07453&0.07228\\
&0.2&0.2611$-$0.2617&0.2606\\
&0.4&0.4461$-$0.4465&0.4454\\
&0.6&0.6288$-$0.6291&0.6288\\
&0.8&0.8126$-$0.8129&0.8113\\
&1&1.001$-$1.001&1.000\\ \hline\hline
\end{tabular}
\label{tab1}
\end{center}
   \caption{A Bermudan put option with XVA (10 exercise dates, expiry $T=0.5,1$) in the CEV-like model for the 2nd-order approximation of the characteristic function, and an LSM comparison.}
\end{table}



\subsection{A numerical example for CVA}\label{sec62}
In this section we validate the accuracy of the method presented in Section \ref{sec52} and compute the CVA in the case of unilateral CCR under the model dynamics given in Section \ref{sec2} with a local jump measure and a
local volatility function with CEV-like dynamics, Gaussian jumps defined by defined as in \eqref{eq:parameters2} and a local default function $\gamma(x)=ce^{\beta x}$. We assume the same parameters as in Section \ref{sec62}, except $r=0.05$ and we take $c = 0.1$ in the default function. In the LSM the number of time steps is taken to be 100 and we simulate $10^5$ paths. In the COS method we take $L = 10$ and $J=100$. Again, due to the state-dependent coefficients in the underlying dynamics we use the approximated characteristic function as derived in Section \ref{sec42} with the second-order approximation, i.e. $\hat \Gamma^{(2)}(t,x;T,\xi)$ and take $\bar x = X_0$. 

\begin{figure}[h]
\begin{center}
\includegraphics[scale = 0.3]{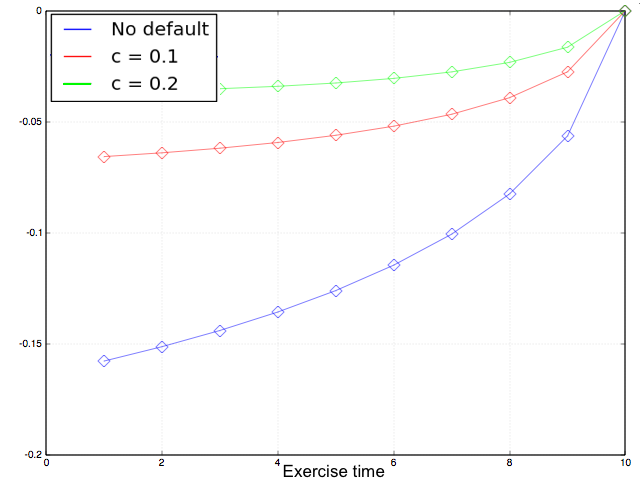}
\end{center}
   \caption{Optimal exercise boundary for a Bermudan put option (10 exercise dates, expiry $T=1$) in the CEV-like model with varying default $c=0,0.1,0.2$.}
\label{fig2}
\end{figure}

The results for the CVA valuation with the FFT-based method and with LSM are presented in Table \ref{tab0}. The CPU time of the LSM is at least 5 times the CPU time of the approximating method, which for $M$ exercise dates is approximately $3\cdot M$ ms, thus more efficient than the computation of the XVA with the method in Section \ref{sec51}. The optimal exercise boundary in Figure \ref{fig2} shows that the exercise region becomes larger when the probability of default increases; this is to be expected: in case of the default probability being greater, the option of exercising early is more valuable and used more often.

\begin{table}[h]
\label{tab0}
\begin{center}
\begin{tabular}{ l|l|l|l }
maturity $T$&strike $K$ &MC CVA&COS CVA\\\hline \hline
0.5&0.6&$4.200\cdot 10^{-4}-4.807\cdot 10^{-4}$&$1.113\cdot 10^{-4}$\\
&0.8&0.001525$-$0.001609&9.869$\cdot 10^{-4}$\\
&1&0.01254$-$0.01273&0.01138\\
&1.2&0.005908$-$0.005931&0.005937\\
&1.4&0.006657$-$0.06758&0.006898\\
&1.6&0.007795$-$0.008008&0.007883\\ \hline \hline
1&0.6&8.673E-04$-$9.574E-04&4.463E-04\\
&0.8&0.005817$-$0.006040&0.003535\\
&1&0.02023$-$0.02054&0.01882\\
&1.2&0.01221$-$0.01222&0.1272\\
&1.4&0.01378$-$0.01391&0.01360\\
&1.6&0.01532$-$0.01502&0.01554\\
\hline\hline
\end{tabular}
\end{center}
   \caption{CVA for a Bermudan put option (10 exercise dates, expiry $T=0.5,1$) in the CEV-like model for the 2nd-order approximation of the characteristic function, and an LSM comparison.}
\end{table}

\section{Conclusion}
In this paper we considered pricing Bermudan derivatives under the presence of XVA, consisting of CVA, DVA, MVA, FVA and KVA. We derived the replicating portfolio with cashflows corresponding to the different rates for different types of lending. This resulted in the PIDE in \eqref{eq:thepde} and its corresponding BSDE \eqref{eq:fundbsde}. We propose to solve the BSDE using a Fourier-cosine method for the resulting conditional expectations and an adjoint expansion method for determining an approximation of the characteristic function of the local L\'evy model in \eqref{eq:hetmodel}. This approach is extended to Bermudan option pricing in Section \ref{sec51}. In Section \ref{sec52} we presented an alternative for computing the CVA term in the case of unilateral collateralization (as is the case when the derivative is an option) without the use of BSDEs. This results in an even more efficient method due to the ability to use the FFT. We verify the accuracy of both methods in Sections \ref{sec61} and \ref{sec62} by comparing it to a LSM and conclude that the method from Section \ref{sec51} is able to achieve a rapid convergence and gives, already for small values of the discretization parameters an accurate result. The alternative method for CVA computation from Section \ref{sec52} is indeed more efficient than the BSDE method for computing just the CVA term. 

\section*{Acknowledgments}
We thank two anonymous referees for the comments and suggestions that have improved the quality of this paper. This research is supported by the European Union in the the context of the H2020 EU Marie Curie Initial Training Network project named WAKEUPCALL. 

\appendix
%

\section{The COS formulae}\label{app1}
Let, as usual, $J$ denote the number of Fourier-cosine coefficients.
Remembering that the expected value $c(t,x)$ in \eqref{eq:bermud} can be rewritten in integral form, we have
\begin{align}
 c(t,x) = e^{-r(t_{m}-t)}\int_\mathbb{R}  v(t_m,y)\Gamma(t,x;t_{m},dy),\qquad t\in[t_{m-1},t_{m}[,
\end{align}
where, $v(t_m,y)$ can be either $u(t_m,y)$ or $\hat u(t_m,y)$.
Then we use the Fourier-cosine expansion to get the approximation:
\begin{align}\label{eq:conti}
 &\hat c(t,x)= e^{-r(t_{m}-t)}\sideset{}{'}\sum_{j=0}^{J-1}
 \textnormal{Re}\left(
 e^{-ij\pi\frac{a}{b-a}}\hat\Gamma\left(t,x;t_{m},\frac{j\pi}{b-a}\right)\right)V_j(t_{m}),\qquad t\in[t_{m-1},t_{m}[\\
 &V_j(t_m)=\frac{2}{b-a}\int_a^b \cos\left(j\pi \frac{y-a}{b-a}\right)\max\{\phi(t_{m},y),c(t_{m},y)\}dy,
\end{align}
with $\phi(t,x)=\left(K-e^{x}\right)^{+}$.

We can recover the coefficients $\left(V_j(t_m)\right)_{j=0,1,...,J-1}$ from
$\left(V_j(t_{m+1})\right)_{j=0,1,...,J-1}$. To this end, we split the integral in the definition
of $V_j(t_m)$ into two parts using the early-exercise point $x_m^*$, which is the point where the
continuation value is equal to the payoff, i.e. $c(t_m,x_m^*)=\phi(t_m,x_m^*)$; this point can easily be found by using the Newton method. Thus, we have
 $$V_j(t_m)=F_j(t_{m},x_m^*)+C_j(t_{m},x_m^*),\qquad m=M-1,M-2,...,1,$$
where
\begin{equation}\label{eq:vcoef}
\begin{split}
 F_j(t_{m},x_m^*)&:=\frac{2}{b-a}\int_a^{x_m^*}\phi(t_m,y)\cos\left(j\pi\frac{y-a}{b-a}\right)dy,\\
 C_j(t_{m},x_m^*)&:=\frac{2}{b-a}\int_{x_m^*}^b c(t_m,y)\cos\left(j\pi\frac{y-a}{b-a}\right)dy,
\end{split}
\end{equation}
and $V_j(t_M) =F_j(t_{M},\log K).$

The coefficients $F_j(t_m,x_m^*)$ can be computed analytically using $x_m^*\leq \log
K$, and by inserting the approximation \eqref{eq:conti} for
the continuation value into the formula for $C_j(t_{m},x_{m}^*)$ have the following coefficients
$\hat C_j$ for $m =M-1,M-2,...,1$:
\begin{align}
 \hat C_j(t_{m},x_m^*) =& \frac{2e^{-r(t_{m+1}-t_{m})}}{b-a}\nonumber\\\label{eq:contin}
 &\cdot\sideset{}{'}\sum_{k=0}^{J-1}V_k(t_{m+1})\int_{x_m^*}^{b}
 \mathrm{Re}\left(e^{-ik\pi\frac{a}{b-a}}\hat\Gamma\left(t_{m},x;t_{m+1},\frac{k\pi}{b-a}\right)\right)
 \cos\left(j\pi\frac{x-a}{b-a}\right)dx.
\end{align}
From \eqref{eq:struc3} we know that the $n$th-order approximation of the characteristic function is
of the form:
\begin{align}
  \hat\Gamma^{(n)}(t_m,x;t_{m+1},\xi)= e^{i\xi x} \sum_{h=0}^n (x-\bar x)^h g_{n,h}(t_m,t_{m+1},\xi),
\end{align}
where the coefficients $g_{n,h}(t,T,\xi)$, with $0\leq k\leq n$, depend only on $t,T$ and $\xi$, but not on $x$. 
\begin{remark}[The defaultable and default-free characteristic functions]
To find $u(t,x)$ we use $$\hat\Gamma^r(t_m,x;t_{m+1},\xi):=e^{i\xi x} \sum_{h=0}^n (x-\bar x)^h g_{n,h}^r(t_m,t_{m+1},\xi),$$ the characteristic function with $\gamma(t,x)=0$. For $\hat u (t,x)$ we use $$\hat\Gamma^d(t_m,x;t_{m+1},\xi):=e^{i\xi x} \sum_{h=0}^n (x-\bar x)^h g_{n,h}^d(t_m,t_{m+1},\xi),$$ where $\gamma(t,x)$ is chosen to be some specified function.
\end{remark} 
Using \eqref{eq:struc3} we can write the Fourier coefficients of the continuation value in vectorized form as:
\begin{align}
\bold{\hat C}(t_{m},x_m^*) =\sum_{h=0}^n e^{-r(t_{m+1}-t_m)}
\mathrm{Re}\left(\bold V(t_{m+1})\mathcal{M}^h(x_m^*,b)\Lambda^h\right),
\end{align}
where $\bold V(t_{m+1})$ is the vector $[V_0(t_{m+1}),...,V_{J-1}(t_{m+1})]^T$ and
$\mathcal{M}^h(x_m^*,b)\Lambda^h$ is a matrix-matrix product with $\mathcal{M}^h$ a matrix
with elements $\{M_{k,j}^h\}_{k,j=0}^{J-1}$ defined as
\begin{align}M_{k,j}^h(x_m^*,b) := \frac{2}{b-a}\int_{x_m^*}^{b}  e^{ij\pi\frac{x-a}{b-a}}(x-\bar x)^h\cos\left(k\pi\frac{x-a}{b-a}\right)dx\label{eq:integraal},\end{align}
and $\Lambda^h$ is a diagonal matrix with elements
 $$g_{n,h}\Big(t_m,t_{m+1},\frac{j\pi}{b-a}\Big),\qquad j=0,\dots,J-1.$$
One can show, see \cite{borovykh}, that the resulting matrix $\mathcal{M}^h$ is a sum of a Hankel and Toeplitz matrix and thus the resulting matrix vector product can be calculated using a FFT.

\bibliographystyle{siam}
\bibliography{Biblio}

\end{document}